\documentstyle[twocolumn,seceq]{jpsj}

\newcommand{\be}{\begin{equation}}
\newcommand{\ee}{\end{equation}}
\newcommand{\bea}{\begin{eqnarray}}
\newcommand{\eea}{\end{eqnarray}}
\newcommand{\lsim}{\raise.35ex\hbox{$<$}\kern-0.75em\lower.5ex\hbox{$\sim$}}
\newcommand{\gsim}{\raise.35ex\hbox{$>$}\kern-0.75em\lower.5ex\hbox{$\sim$}}

\def\runtitle{Dynamical Properties of the One-Dimensional
  Supersymmetric {\it t-J} Model}
\def\runauthor{Yasuhiro {\sc Saiga} and Yoshio {\sc Kuramoto}}

\title
{
Dynamical Properties of the One-Dimensional Supersymmetric \\
\mbox{\boldmath $t$}-\mbox{\boldmath $J$} Model:
A View from Elementary Excitations
}

\author
{ 
Yasuhiro {\sc Saiga}\footnote{E-mail: saiga@ginnan.issp.u-tokyo.ac.jp}
and Yoshio {\sc Kuramoto}$^1$
}

\inst
{
Institute for Solid State Physics, University of Tokyo, Roppongi
7-22-1, Minato-ku, Tokyo 106-8666 \\
$^1$Department of Physics, Tohoku University, Sendai 980-8578
}

\recdate
{\hspace{2cm}}

\abst
{
Dynamical properties, such as dynamical spin and charge structure
factors and single-particle spectral functions, are studied for
the one-dimensional supersymmetric {\it t-J} model with inverse-square
interaction.
Exact diagonalization and the recursion method are used for finite
systems up to 16 sites.
A simple rule is proposed to understand the supermultiplet structure
in the excitation spectrum.
Numerical calculations show that the dynamical spin structure factor
is independent of the electron density in the region where only two
spinons contribute.
In the electron removal spectrum from half-filling, close correspondence 
is found between the dominant contribution from three-spinon plus one-holon states in the {\it t-J} model 
and the two-spinon excitations in the Haldane-Shastry model.
Comparison is made with dynamics of the nearest-neighbor supersymmetric {\it t-J} model.
}

\kword
{
dynamical spin structure factor, dynamical charge structure factor,
single-particle spectral function, {\it t-J} model, supersymmetry,
inverse-square interaction, nearest-neighbor interaction, exact
diagonalization, recursion method
}

\begin{document}
\sloppy
\maketitle



\section{Introduction}

One-dimensional (1D) electron systems show strong quantum fluctuations, and have been investigated 
intensively both theoretically and experimentally.
Most of real quasi-1D systems have localized electrons and are regarded as spin systems.
Recently some systems with doped holes have also been studied experimentally.
For example, Yb$_4$As$_3$ has a small number of carriers, 
and shows a heavy-fermion behavior in resistivity, magnetic susceptibility and specific 
heat.~\cite{Ochiai}
The spin dynamics in Yb$_4$As$_3$ probed at low temperature suggests that the 
nature of low-energy spin excitations
is similar to that of the 1D antiferromagnetic Heisenberg model.~\cite{Kohgi97,Kohgi99}
On the other hand, a compound Sr$_{14}$Cu$_{24}$O$_{41}$ consists of two-leg Cu$_2$O$_3$ ladders and 
edge-sharing CuO$_2$ chains.
The number of holes in the chain is estimated to be 60\% of the number
of Cu ions.~\cite{Carter}
One may ask how the existence of holes affects dynamical properties
 of quasi-1D systems.
In the presence of both spin and charge degrees of freedom, 
one can expect much richer physics as compared with spin-only systems.
\par

Recent angle-resolved photoemission measurements have revealed the
existence of spin-charge separation in quasi-1D systems.
In Mott insulators such as SrCuO$_2$~\cite{Kim96,Kim97} and
Sr$_2$CuO$_3$,~\cite{Fujisawa} angle-resolved photoemission data show
two distinct dispersions in the energy-momentum space.
One is assigned to the holon band, and the other is regarded as the
spinon band.
This is an indication of the spin-charge separation.
More recently, an angle-resolved photoemission study on
PrBa$_2$Cu$_3$O$_7$ has been performed.~\cite{Mizokawa}
Signals from the doped CuO$_3$ chain reveal the spinon and holon bands
in the range of momenta from 0 to $\pi/4$ in units of the reciprocal
lattice parameter.
This indicates that the CuO$_3$ chain is nearly quarter-filled.
In order to investigate theoretically dynamics of these systems with holes,
one can take the 1D {\it t-J} model~\cite{THM,ZKMS} as the simplest model.
To understand the photoemission data, in particular, study of hole
propagators in the {\it t-J} model is necessary.~\cite{SP,MTY,TM,PHMS96,PHMS97,FHPMD,EO}
\par

For the {\it t-J} model with nearest-neighbor interaction and
small $J/t$, which is ordinarily adopted, dynamical correlation functions are rather intricate.
This complexity comes from the fact that Landau quasi-particles are no
longer well-defined elementary excitations.
It is known that elementary excitations in 1D {\it t-J} model obey
fractional statistics.
In this context, the long-range supersymmetric {\it t-J}
model~\cite{KY} is suited for extracting the essential dynamical
properties since the elementary excitations do not interact except for
the generalized exclusion principle.~\cite{Haldane91} 
For the long-range {\it t-J} model, the excitation contents are
composed of {\it spinons}, {\it holons} and {\it antiholons}; the
number of elementary excitations contributing
to dynamics is {\it finite} due to the high symmetry of the model.~\cite{HH}
Ha and Haldane proposed regions of nonvanishing spectral weight in
the energy-momentum space, which is called the ^^ ^^ compact
support", for various dynamical correlation functions.~\cite{HH}
Recently, exact results for the spectral weight itself have been
partially derived.~\cite{Kato,AYSK}
However, full analytic solution of the dynamical correlation function is
not yet found.
\par

In this paper, we clarify the dynamical properties of the 1D
supersymmetric {\it t-J} model with long-range interaction. 
We calculate the dynamical spin and charge structure factors and the
single-particle spectral functions for systems 
up to 16 sites
with various hole numbers via exact diagonalization and the
recursion method.
Using the asymptotic Bethe-ansatz equations and the skew Young
diagram, we reexamine the excitation contents proposed by Ha and
Haldane.~\cite{HH}
The relationship between the calculated dynamical quantities and the
excitation contents is investigated.
The results of dynamical quantities are compared with those for the
nearest-neighbor supersymmetric {\it t-J} model,~\cite{SutherlandtJ}
which is solvable by the Bethe ansatz.
\par

The remainder of this paper is arranged as follows.
In $\S$2 we consider two boundary conditions for the long-range {\it
  t-J} model: the periodic boundary condition and the antiperiodic
one.
Particularly, the antiperiodic boundary condition is nontrivial for
the long-range {\it t-J} model.
In $\S$3 we show results of various dynamical quantities and interpret
them in terms of elementary excitations.
Finally we give the summary of this study and discuss the results in
$\S$4.
\par


\section{Model and Boundary Conditions} \label{MBC}

We consider the 1D {\it t-J} model
\bea
  {\cal H} &=& \sum_{i<j}^{N} \sum_{\ell = - \infty}^{\infty} \Biggl[ - t_{ij \ell} \sum_{\sigma} \left( 
\tilde{c}^{\dagger}_{i \sigma} \tilde{c}_{j + \ell N, \sigma} + {\rm H. c.} \right) \nonumber \\
           & & + J_{ij \ell} \left( \mbox{\boldmath $S$}_i \cdot \mbox{\boldmath $S$}_{j + \ell N} - \frac{1}{4} n_i 
n_{j + \ell N} \right) \Biggr],\label{tJHamiltonian}
\eea
with even number $N$ of sites and average electron number $\bar{n}$ per site.
Here $\tilde{c}_{i \sigma} = c_{i \sigma} (1 - n_{i,- \sigma})$ is the annihilation operator of an electron with 
spin $\sigma$ at site $i$ with the constraint of no double occupation.
The summation with respect to $\ell$ has been introduced to treat the
system in a ring geometry.~\cite{Sutherland,FK}
Two types of the interaction are considered:

\noindent (i) the nearest-neighbor (NN) type, i.e., $t_{ij \ell} = t \delta_{j,i+1} \delta_{\ell,0}$;

\noindent (ii) the inverse-square (IS) type, i.e., $t_{ij \ell} = t / (x_i - x_j - \ell N)^2$.

\noindent The Hamiltonian in the cases (i) and (ii) are written as ${\cal H}_{\rm NN}$ and ${\cal H}_{\rm IS}$, 
respectively.
\par

For the NN {\it t-J} model, it is conventional to choose the periodic
boundary condition $c_{j+N,\sigma} = c_{j \sigma}$ for $N_{\rm e} =
4m + 2$ with $N_{\rm e}$ being the 
electron number and $m$ an integer,~\cite{OLSA} 
and the antiperiodic boundary condition $c_{j+N,\sigma} = -
c_{j \sigma}$ for $N_{\rm e} = 4m$.
This leads to the nondegenerate ground state which is a singlet with zero total momentum.
The Hamiltonian is written out as follows:
\bea
  {\cal H}_{\rm NN} &=& \sum_{i=1}^{N} \Biggl[ - t \sum_{\sigma} \left( \tilde{c}^{\dagger}_{i \sigma} 
\tilde{c}_{i+1, \sigma} + {\rm H. c.} \right) \nonumber \\
                    & & + J \left( \mbox{\boldmath $S$}_i \cdot \mbox{\boldmath $S$}_{i+1} - \frac{1}{4} n_i n_{i+1} 
\right) \Biggr],\label{NNtJHamiltonian}
\eea
where $\tilde{c}_{N+1,\sigma} = \tilde{c}_{1 \sigma}$ for $N_{\rm e} = 4m + 2$, and $\tilde{c}_{N+1,\sigma} = - 
\tilde{c}_{1 \sigma}$ for $N_{\rm e} = 4m$; in addition, $\mbox{\boldmath $S$}_{N+1} = \mbox{\boldmath 
$S$}_1$ and $n_{N+1} = n_1$ in any case.
\par

For the IS {\it t-J} model, on the other hand, the effect of the boundary condition is nontrivial.
When the periodic boundary condition $c_{j+\ell N,\sigma} = c_{j \sigma}$ are imposed, both transfer energy 
and exchange energy become the sine-inverse-square type given by~\cite{KY}
\bea
  t_{ij}^{(\rm p)} &=& t D(x_i - x_j)^{-2},\label{tijpbc} \\
  J_{ij} &=& J D(x_i - x_j)^{-2},\label{Jijpbc}
\eea
where $D(x_i - x_j) = (N/\pi) \sin [\pi (x_i - x_j)/N]$.
Under the antiperiodic boundary condition $c_{j+\ell N,\sigma} = (-1)^\ell c_{j \sigma}$,
on the other hand, the transfer energy $t_{ij}^{(\rm a)}$ is given by
\bea
  t_{ij}^{(\rm a)} &=& t \sum_{\ell = -\infty}^\infty \frac{(-1)^\ell}{(x_i-x_j-\ell N)^2} \nonumber \\
                   &=& t D(x_i - x_j)^{-2}  \cos [\pi (x_i-x_j)/N].
\eea
This corresponds to the special case of the twisted boundary
condition.~\cite{FK,LW}

We want to consider the spin and charge dynamics from the nondegenerate ground state.
Hence for $N_{\rm e} = 4m + 2$  we choose the periodic boundary condition, 
and for $N_{\rm e} = 4m$ we impose the antiperiodic one.
Consequently, we consider the following Hamiltonian as the {\it t-J} model with IS interaction:
\bea
  {\cal H}_{\rm IS} &=& \sum_{i<j}^{N} \Biggl[ - \tilde{t}_{ij} \sum_{\sigma} \left( \tilde{c}^{\dagger}_{i \sigma} 
\tilde{c}_{j \sigma} + {\rm H. c.} \right) \nonumber \\
                    & & + J_{ij} \left( \mbox{\boldmath $S$}_i \cdot \mbox{\boldmath $S$}_j - \frac{1}{4} n_i n_j 
\right) \Biggr],\label{IStJHamiltonian}
\eea
where $J_{ij}$ is given by eq.\ (\ref{Jijpbc}), and
\be
  \tilde{t}_{ij} = \left\{
   \begin{array}{ll}
    t_{ij}^{(\rm p)} \quad &\mbox{for $N_{\rm e} = 4m + 2$},\\
    t_{ij}^{(\rm a)} \quad &\mbox{for $N_{\rm e} = 4m$}.
   \end{array}\right.\label{tildetij}
\ee
When we calculate the single-particle spectral functions from a
less-than-half-filled ground state, we start from the initial state with
$N_{\rm e} = 4m + 2$ and therefore use $\tilde{t}_{ij} = t_{ij}^{(\rm p)}$.
We always take the supersymmetric case $t = J/2$ in the following  calculation.
\par

The dynamical quantities we consider in this paper are the followings:
the dynamical spin structure factor $S(q,\omega)$, the dynamical charge structure factor $N(q,\omega)$, 
and the single-particle spectral function $A(k,\omega)$.
These quantities are defined as follows:
\bea
  S^{zz}(q,\omega) &=& \sum_\nu \left| \langle \nu |s_q^z| 0 \rangle \right|^2 \delta(\omega -E_\nu +E_0), 
\\
  N(q,\omega) &=& \sum_\nu \left| \langle \nu |n_q| 0 \rangle \right|^2 \delta(\omega -E_\nu +E_0), \\
  A^-(k,\omega) &=& \sum_\nu \left| \langle \nu; N_{\rm e}-1 |\tilde{c}_{k \sigma}| 0; N_{\rm e} \rangle 
\right|^2 \nonumber \\
                & & \times \delta(\omega + E_\nu (N_{\rm e}-1) - E_0 (N_{\rm e})), \label{A-komega} \\
  A^+(k,\omega) &=& \sum_\nu \left| \langle \nu; N_{\rm e}+1 |\tilde{c}_{k \sigma}^\dagger| 0; N_{\rm e} 
\rangle \right|^2 \nonumber \\
                & & \times \delta(\omega - E_\nu (N_{\rm e}+1) + E_0 (N_{\rm e})),
\eea
where $s_q^z = N^{-1/2} \sum_\ell s_\ell^z {\rm e}^{- {\rm i} q \ell}$, $n_q = N^{-1/2} \sum_\ell (n_\ell - 
\bar{n}) {\rm e}^{- {\rm i} q \ell}$, 
and $\tilde{c}_{k \sigma} = N^{-1/2} \sum_\ell \tilde{c}_{\ell \sigma} {\rm e}^{- {\rm i} k \ell}$;
$q$ (or $k$) and $\omega$ represent the momentum transfer and the excitation energy, respectively.
$| \nu \rangle$ denotes an eigenstate of ${\cal H}$ with energy $E_\nu$ with $E_0$ being the ground-state 
energy.
These dynamical quantities can be written in the form of a continued fraction.~\cite{GB}
We truncate the continued fraction after 100-200 iterations.
In our calculation, poles and intensities have 
accuracies of 7 digits and 4-5 digits, 
respectively.~\cite{comment}
\par


\section{Dynamical Properties and Elementary Excitations} \label{DPEE}

\subsection{Ha-Haldane's support}

Before presenting our results on dynamical properties, we briefly review the compact supports for various 
correlation functions for the IS supersymmetric {\it t-J} model.
The compact support means a finite region where the spectral function is nonzero.
To draw the support, one needs three pieces of information:~\cite{HH}
dispersion relations of related elementary excitations, excitation
contents of the intermediate states for each correlation function, and
selection rules.
The dispersion relations for the right(left)-going spinon (s$_{{\rm R}({\rm L})}$), holon (h$_{{\rm R}({\rm L})}$) 
and antiholon ($\bar{\rm h}$) in the thermodynamic limit are given by
\bea
  & & \epsilon_{{\rm s}_{\rm R(L)}}/t = - p (p \mp v_{\rm s}^0),\label{spinondispersion} \\
  & & \epsilon_{{\rm h}_{\rm R(L)}}/t = p (p \pm v_{\rm c}^0),\label{holondispersion} \\
  & & \epsilon_{\bar{\rm h}}/t = \frac{1}{2} \left[ \left( v_{\rm c}^0 \right)^2 - p^2 
\right],\label{antiholondispersion}
\eea
where $v_{\rm s}^0 = \pi$ and $v_{\rm c}^0 = \pi (1 - \bar{n})$.
The range of momentum permissible for each elementary excitation 
depends on the electron density $\bar{n}$.
The right (left) spinons and holons are allowed in $0 \le p \le k_{\rm 
  F}$ ($- k_{\rm F} \le p \le 0$), where $k_{\rm F} = \pi \bar{n}/2$.
The antiholons propagate in the region $- v_{\rm c}^0 \le p \le v_{\rm c}^0$.
The excitation contents, which mean all possible states excited
from the ground state by a local operator, are determined with the
help of the numerical results.
Finally, the selection rules are found empirically, also with use of the numerical information.

\subsection{Derivation of excitation contents} \label{Econtent}

Although exact expression of the matrix element, or the form factor, is not yet obtained, one can 
determine the excitation contents by comparing numerically obtained levels
with those in the asymptotic Bethe-ansatz equations.
In this subsection, we explain how to derive the excitation contents corresponding to each energy level.
\par

Let us take the completely up-polarized state as the reference state and consider  
$M+Q$ pseudo-particles which represent $M$ electrons with spin down and $Q$ holes.
Note that these pseudo-particles are not spinons and holons.
The asymptotic Bethe ansatz leads to the following equation:~\cite{Kawakami,WLC}
\be
  \frac{E}{t} = \frac{\pi^2}{3} Q \left( 1 - \frac{1}{N^2} \right) + \sum_{\mu = 1}^{M+Q} \bar{\epsilon}(p_{\mu}),
\label{EnergyABA}
\ee
where $\bar{\epsilon}(p) = p (p - 2 \pi)/2$.
The pseudomomenta $p_{\mu}$ are determined by the following equations:
\bea
  p_{\mu} N &=& 2 \pi J_{\mu} - \pi \sum_{i=1}^Q {\rm sgn} (p_{\mu} - q_i) \nonumber \\
            & & + \pi \sum_{\nu = 1}^{M+Q} {\rm sgn} (p_{\mu} -
            p_{\nu}),\label{momentumABA1} \\
  2 \pi I_i &=& \pi \sum_{\mu = 1}^{M+Q} {\rm sgn} (q_i - p_{\mu}),\label{momentumABA2}
\eea
where $J_{\mu} \in [(M+1)/2, N-(M+1)/2]$ for $\mu=1, 2, \cdots, M+Q$,
and $I_i \in [-(M+Q)/2, (M+Q)/2]$ for $i=1, 2, \cdots, Q$.
The quantum numbers $J_\mu$ and $I_i$ are integers or half-integers, and are arranged in the ascending 
order.
We assume that all $p_\mu$ and $q_i$ are distinct from each other.
For given $p_\mu$, we specify $j(\mu)$ such that
$q_{j(\mu)} < p_\mu < q_{j(\mu)+1}$ in the case of $0 < j (\mu) < Q$, 
$j(\mu)=0$ if $p_\mu < q_1$ and $j(\mu)=Q$ if $q_Q < p_\mu$.
Then we obtain 
\be
  \frac{N}{2 \pi} p_\mu = J_\mu + \mu - \frac{M+1}{2} -  j(\mu),
\label{ABAeqsolution}
\ee
from eqs.\ (\ref{momentumABA1}) and (\ref{momentumABA2}).
Note that $\tilde{p}_\mu \equiv N p_\mu/(2 \pi)$ is always integral.

Identification of excitation contents proceeds as follows: 
First, we give arbitrary configurations of $J_{\mu}$ and $I_i$ in the
subspace with fixed $M$ and $Q$ and obtain a set of
pseudomomenta $\{ p_{\mu} \}$.
Second, using the set $\{ p_{\mu} \}$, we find the total energy $E$,
the total momentum $P = \sum_{\mu = 1}^{M+Q} p_{\mu}$ (mod $2 \pi$),
and the sequence of occupied state (represented by 1) and empty state
(represented by 0) in the pseudomomentum space.
This sequence is referred to as the motif.
The first and last entries of the motif are always set to 0. 
We note that $\bar{\epsilon}( p_\mu )$ takes the maximum value $(=0)$ for $p_\mu = 0$ and $2 \pi$.  
Finally, the sequence (i.e., the motif) is assigned to an eigenstate
with the same energy and momentum obtained by exact
diagonalization.
Let us give an example.
For $(N,Q,M)=(4,2,0)$, permissible values of $J_\mu$ are 1/2, 3/2, 5/2 
and 7/2, while those of $I_i$ are $-1, 0$ and $1$.
We consider the case of $\{ J_\mu \} = \{ 3/2, 7/2 \}$ and $\{ I_i 
\}= \{ -1, 0 \}$.
Then we obtain $\{ \tilde{p}_\mu \} = \{ 1, 3
\}$, so that the motif becomes 01010.
\par

The IS supersymmetric {\it t-J} model has the supermultiplet
structure, i.e., enormous degeneracies in the excitation
spectrum.~\cite{WLC} 
However, the motif itself does not provide us information on the degeneracy.
To extract the information, the skew Young diagram
method~\cite{KKN} is very useful.
The motif and the skew Young diagram have a one-to-one correspondence.
The correspondence goes as follows:
To begin with, we write the first box. 
Then we read a motif from the left, neglecting the first and last
entries of the motif;
if we encounter 1 (0), we add a box over (on the right of) the last box.
We now propose a simple rule to reproduce the supermultiplet structure of the IS {\it t-J} model by 
generalizing the known rule for the SU($m$) symmetry.~\cite{KKN}
Namely one puts an index for the internal degrees of freedom on each box of a skew
Young diagram.
In the present case we represent an up-spin by 1, a down-spin by 2, and a hole by $\circ$.
The order of them is defined as $1 < \circ < 2$.
We introduce the following rule:

\noindent [i] $a < b$ if $b$ is lower-adjacent to $a$;
however, consecutive $\circ$'s are allowed in a column.
Note that the numbers 1 and 2 can appear at most once in a column.

\noindent [ii] $a \ge b$ if $b$ is left-adjacent to $a$;
however, consecutive $\circ$'s are forbidden in a row.
\par

For the 4-site IS {\it t-J} model with the fixed number of
holes, the degeneracy based on this rule is consistent with the numerical result for all eigenvalues.
We note that this rule enables us to decompose the degeneracy for each number of holes.
For instance, the motif 01010 for $N=4$ has the energy 
$E/t = \pi^2 (5Q - 12)/16$ and the momentum $P = 0$.
After translating this motif into the skew Young diagram, we find
that the number of ways to put 1, 2 or $\circ$ on each box is
12, according to [i] and [ii] above (see Fig.\ \ref{fig.SYD}).
The 12-fold degeneracy is decomposed as follows:
nondegenerate for $Q=0$, 4-fold degenerate for $Q=1$, 5-fold for $Q=2$, and 2-fold
for $Q=3$.
These results agree with those via numerical diagonalization: 
$(S_{\rm tot} = 0)^1$ for $Q=0$, $(S_{\rm tot} = 1/2)^2$ for $Q=1$,
$(S_{\rm tot} = 0)^2 \oplus (S_{\rm tot} = 1)^1$ for $Q=2$, and
$(S_{\rm tot} = 1/2)^1$ for $Q=3$. 
\par

We note that the degeneracy in the spectrum has also been studied for the SU($m$)~\cite{KKN} and 
SU($m|n$)  supersymmetric~\cite{HB} Polychronakos spin chains, which are a variant of the $1/r^2$-type 
model.
The SU($1|2$) specialization of the latter chain has the same supersymmetric Yangian symmetry as the IS 
{\it t-J} model.
Thus one may expect that the supermultiplet structures
in the spectrum are the same for both models.
We have checked the following for $N=4$: 
If the first term of eq.\ (\ref{EnergyABA}) is subtracted from an energy level by 
exact diagonalization of the {\it t-J} model, the degeneracy of the 
energy level is consistent with that for the SU($1|2$) supersymmetric
Polychronakos spin chain.\cite{HB}
The subtraction is necessary since the first term of eq.\ (\ref{EnergyABA}) breaks the global SU($1|2$) 
supersymmetry.
\par

Now, we can extract the excitation contents from the motif.
According to Ha and Haldane~\cite{HH}, a spinon (with spin 1/2 and no charge) is represented by two 
consecutive zeros, and a holon (with charge $- {\rm e}$ and no spin) by two consecutive ones.
For less-than-half-filling, the motif of the ground state with no magnetization is 
01010...10$|$111...111$|$01...01010, which has ($N+1$) sequence.
The central region [($Q+1$) consecutive ones] is regarded as a 
``pseudo Fermi sea (PFS)".
A hole in the PFS (i.e., zero in the center region) is called an
antiholon (with charge $+ 2{\rm e}$ and no spin).
Spinons and holons are excited in the left and/or right region.
For example, a motif 00110$|$1111101$|$10100 should read (s$_{\rm L}$, h$_{\rm L}$) + $\bar{{\rm h}}$ + 
(h$_{\rm R}$, s$_{\rm R}$).
\par

When we derive the excitation contents in this way, however, exceptional situations occur in some cases.
For motifs such as 01010$|$1111101$|$01100 and 01010$|$1111110$|$11010, which are actually present 
for 16 sites with 6 holes, excitation contents according to ref.\ 20 become $\bar{{\rm h}}$ + (h$_{\rm 
R}$, s$_{\rm R}$) [case (a)] and $\bar{{\rm h}}$ + h$_{\rm R}$ [case (b)], respectively.
These contents do not fulfill the charge and/or spin conservations, and the latter does not fulfill the 
semionic statistics of holons.
This suggests that the excitation contents by Ha and Haldane~\cite{HH} do not always hold true for finite 
systems.
In contrast, without holes, i.e., for the Haldane-Shastry (HS) spin
model~\cite{Haldane88,Shastry88}, we can always read two spinons from the motif for all states that 
contribute to $S(q,\omega)$ even for finite systems ($N \le 16$).
\par

Excitation contents for a finite system 
can be extracted also by
the skew Young diagram.
We first determine the PFS of the diagram from the corresponding motif.
Note that the PFS consists of $Q$ boxes.
We regard $\circ$ outside the PFS as a holon, and a row of a 1-2 pair
in the PFS as an antiholon.
Single 1 or 2 in a column, both of which do not make a pair, is
regarded as a spinon;
it does not matter whether the column includes $\circ$ or not.
Then, in the case (a), some of the diagrams with the 
entries of 1, 2 and
$\circ$ can read (s$_{\rm L}$, h$_{\rm L}$) + $\bar{{\rm h}}$ +
(h$_{\rm R}$, s$_{\rm R})$.
One of them is shown in Fig.\ \ref{fig.SYDab}(a).
In the case (b), on the other hand, we find from Fig.\
\ref{fig.SYDab}(b) that only 1 of a 1-2 pair in a row is present in
the PFS.
We shall call 
such a part of the 1-2 pair in the PFS a ^^ ^^ half-antiholon
($\bar{\rm h}^\ast$)''.
This quasi-particle is assumed to have charge $+ {\rm e}$, no spin and
semionic statistics.
Thereby the diagram in the case (b) reads $\bar{\rm h}^\ast$ + h$_{\rm
  R}$, which satisfies both charge and spin conservations and the
statistical feature.
In the case where an antiholon in the Ha-Haldane's excitation contents 
is replaced by 
a half-antiholon, a holon should be simultaneously
subtracted from the contents.
As the number of sites increases with hole density 
kept constant,
the effect of the edges in the PFS 
should become negligible.
This means that a half-antiholon has zero energy.
In fact, for $N(q,\omega)$, the energy levels with contents $\bar{\rm
  h}^\ast$ + h$_{\rm R}$ are almost along the holon dispersion for $q \le
k_{\rm F}$ in the compact support.
Thus only Ha-Haldane's contents 
should survive in the thermodynamic limit.
\par

We notice that the spinons and holons identified above do not satisfy
the statistical matrix deduced from 
a microscopic derivation of the exact thermodynamics~\cite{KK} at least for $N=4$.
Nevertheless, these quasi-particles obey the semionic statistics and
exhaust the Hilbert space for the {\it t-J} model.
This reflects 
the fact that a different set of quasi-particle picture is available
in spanning the complete Hilbert space.
\par

For a dynamical correlation function of the finite-size IS {\it t-J}
model, an energy level with the Ha-Haldane's excitation contents
has not always finite intensity.
For example, let us take an eigenvalue with $(q,\omega) = (\pi,
1.09662t)$ for $(N, Q, M) = (12, 2, 5)$.
This eigenvalue has 4-fold degeneracy: $(S_{\rm tot} = 0)^1 \oplus
(S_{\rm tot} = 1)^1$, and its motif 00101$|$101$|$10100 leads to 
(s$_{\rm L}$, h$_{\rm L}$) + $\bar{{\rm h}}$ + (h$_{\rm R}$, s$_{\rm
  R}$).
Nevertheless, there is finite intensity for $S(q,\omega)$ but no
intensity for $N(q,\omega)$.
This implies existence of additional selection rules for $N(q,\omega)$.
\par

\subsection{$S(q,\omega)$: dynamical spin structure factor}

We address to the spectral weight itself for various correlation
functions in the remaining part of this section.
Let us begin with the results of $S(q,\omega)$ for the IS {\it t-J} model.
Figures \ref{fig.IStjsqw}(a) and \ref{fig.IStjsqw}(b) show the results 
in the 16-site chain for electron densities $\bar{n}=0.875$ (14 
electrons) and $0.5$ (8 electrons), respectively.
The solid lines show dispersion relations of elementary excitations for the compact support.
These figures reveal clearly the compactness of the support for the excitation at arbitrary filling; there is no 
intensity outside the outermost dispersion lines.
For two holes, the main intensity is in the low-energy region.
As the hole density becomes larger, the peak frequency at each $q$ seems to approach the antiholon 
dispersion which runs over $k_{\rm F} \le q \le \pi$.
For $\bar{n} = 0.5$ in Fig.\ \ref{fig.IStjsqw}(b), poles with dominant
intensity are nearly along the antiholon dispersion over 
$0.25 \le q/\pi \le 1$.

Let us look at poles and intensities more carefully.
A surprising feature is that poles and intensities in the two-spinon region agree with those for the HS 
model within the numerical accuracy.
We have confirmed this fact for $N \le 16$ with various $\bar{n}$.
It leads to the following conjecture:
{\it independent of $\bar{n}$, the exact expression of $S(q,\omega)$ in the 2s$_{\rm R}$ and 2s$_{\rm L}$ 
regions is identical with that for the HS model},~\cite{HZ} i.e.,
\be
  S(q,\omega) = 1 \Big/ \left\{ 4 \left[ (\omega - \omega_{{\rm L}-}(q)) (\omega - \omega_{{\rm L}+}(q)) 
\right]^{1/2} \right\}, \label{exactSqomega}
\ee
where $\omega_{{\rm L}-}(q) = t q (\pi - q)$ and $\omega_{{\rm L}+}(q) = t (q - \pi) (2\pi - q)$.
This strong spin-charge separation in $S(q,\omega)$ should be a
manifestation of the supersymmetric Yangian symmetry of the present model.
Note that in the NN case such strong separation does not occur.

In the region including holons and antiholons, on the other hand, the $\omega$-dependence of 
$S(q,\omega)$ is more complicated.
Possible reasons for this complication are:
\begin{itemize}
\item For a finite system, the intermediate states of $S(q,\omega)$
  are not always (s$_{\rm L}$, h$_{\rm L}$) + $\bar{{\rm h}}$ +
  (h$_{\rm R}$, s$_{\rm R}$) states, in contrast with the claim by Ha
  and Haldane.~\cite{HH}
  As mentioned in $\S$\ref{Econtent}, the intermediate states include
  the states with contents (s$_{\rm L}$, h$_{\rm L}$) + $\bar{\rm
    h}^\ast$ + s$_{\rm R}$ and the mirror ones (L $\leftrightarrow$ R).
\item It is likely that the intensities of states with different excitation contents merge to a single pole.
\end{itemize}

Now we consider the static spin structure factor $S(q)$, which is obtained by integration of $S(q,\omega)$ 
over $\omega$.
Independent of the IS {\it t-J} model, Gebhard and Vollhardt~\cite{GV} calculated $S(q)$ associated with the 
Gutzwiller wave function~\cite{Gutzwiller} as follows:
\be
  S(q) = \left\{
   \begin{array}{ll}
    - \displaystyle{\frac{1}{4}} \ln \left( 1 - \frac{q}{\pi} \right), \quad &\mbox{for $0 \le q \le 2 k_{\rm F}$},\\
    - \displaystyle{\frac{1}{4}} \ln ( 1 - \bar{n} ), \quad &\mbox{for $2 k_{\rm F} \le q \le \pi$}.
   \end{array}\right.\label{Sq}
\ee
It was then demonstrated that the ground-state wave function of
the IS {\it t-J} model is identical to the Gutzwiller one.~\cite{KY}
Thus eq.\ (\ref{Sq}) turns out to be the exact expression of $S(q)$ for the IS {\it t-J} model.
For $0 \le q \le k_{\rm F}$, integration of eq.\ (\ref{exactSqomega}) over $\omega$ reproduces the former 
of eq.\ (\ref{Sq}).
This supports our conjecture on $S(q,\omega)$ in the two-spinon region .
In Fig.\ \ref{fig.IStjsq} we show the results of $S(q)$ for $N = 16$ for various fillings.

To compare with results for the IS model, we show the counterparts for the NN model.
Figures \ref{fig.NNtjsqw}(a) and \ref{fig.NNtjsqw}(b) show the results of $S(q,\omega)$ for two holes and 
eight holes in the 16-site chain, respectively.
For two holes, the global features are similar to those for the IS model.
However, there are two remarkable differences: 

[A] Contribution of higher-order spinons, holons and antiholons to
$S(q,\omega)$ spreads over the high-energy region. 

[B] Even below the spinon dispersion, there are appreciable intensities for $0 \le q \le k_{\rm F}$. 
This corresponds to the lower holon dispersion in the NN
model.~\cite{Schlottmann}

For eight holes, i.e., quarter-filling [see Fig.\
\ref{fig.NNtjsqw}(b)], the region with main intensity in the
$(q,\omega)$-plane is rather different from the region with finite
intensity for the IS type.
This is attributed to the change of elementary excitations with increasing hole density in the NN 
model.~\cite{BBO}

\subsection{$N(q,\omega)$: dynamical charge structure factor}

We discuss the results of $N(q,\omega)$ for the IS {\it t-J} model.
Figures \ref{fig.IStjnqw}(a) and \ref{fig.IStjnqw}(b) show the results for 16 sites and various electron 
densities.
All poles with finite intensity are inside the corresponding support 
which is indicated with the solid lines in the figures.
For two holes in Fig.\ \ref{fig.IStjnqw}(a), a typical example of lightly hole-doped case, the following features 
are observed:
\begin{itemize}
\item The peak frequency of $N(q,\omega)$ for each $q$ is at the upper edge of the continuum.
\item The intensity of lowest-energy at $q = 2 \pi - 4 k_{\rm F} =
  \pi/4$ is stronger than that at $q = 2 k_{\rm F} = 7 \pi/8$ for
  $\bar{n} = 0.875$.
\end{itemize}
As the hole density is increased, these features disappear gradually.

According to Ha and Haldane,~\cite{HH} the intermediate states of $N(q,\omega)$
consist of states with contents $\bar{{\rm h}}$ + 2h$_{\rm R}$, the
mirror ones, and the ones with contents (s$_{\rm L}$, h$_{\rm L}$) +
$\bar{{\rm h}}$ + (h$_{\rm R}$, s$_{\rm R}$).
We look at the intensity in the pure $\bar{{\rm h}}$ + 2h$_{\rm R}$ region, which does not overlap with the 
region including spinons.
It is found that the intensity in the region increases
monotonically with $\omega$.
We note that $N(q,\omega)$ in the region $q \le k_{\rm F}$ has recently been computed 
analytically.~\cite{AYSK}

Next, we discuss the static charge structure factor $N(q)$.
Like $S(q)$, the exact expression of $N(q)$ is derived by Gebhard and Vollhardt.~\cite{GV}
We define the following functions:~\cite{YO}
\bea
  & & N_1 (x) = x - \frac{x}{2} \ln \frac{( 1 - \bar{n} + x )}{( 1 - \bar{n} )},\\
  & & N_2 (x) = x - \frac{x}{2} \ln \frac{( 1 + \bar{n} - x )}{( 1 - \bar{n} )} \nonumber \\
  & & \hspace{1.3cm} + \ln ( 1 + \bar{n} - x ),\\
  & & N_3 (x) = 2 - 2 \bar{n} + \frac{x}{2} \ln \frac{( \bar{n} - 1 + x )}{( 1 - \bar{n} + x )},\\
  & & N_4 = 2 \bar{n} + \ln ( 1 - \bar{n} ),\\
  & & N_5 (x) = 2 - 2 \bar{n} + \frac{x}{2} \ln \frac{( \bar{n} - 1 + x )}{( 1 + \bar{n} - x )} \nonumber \\
  & & \hspace{1.3cm} + \ln ( 1 + \bar{n} - x ).
\eea
With use of these equations, when $0 < \bar{n} \le 1/2$, one has
\be
  N(q) = \left\{
   \begin{array}{ll}
    N_1 (q/\pi), \quad &\mbox{for $0 \le q \le 2 k_{\rm F}$},\\
    N_2 (q/\pi), \quad &\mbox{for $2 k_{\rm F} \le q \le 4 k_{\rm F}$},\\
    N_4, \quad &\mbox{for $4 k_{\rm F} \le q \le \pi$}.
   \end{array}\right.\label{Nqregion1}
\ee
For $1/2 \le \bar{n} \le 2/3$, one has
\be
  N(q) = \left\{
   \begin{array}{ll}
    N_1 (q/\pi), \quad &\mbox{for $0 \le q \le 2 k_{\rm F}$},\\
    N_2 (q/\pi), \quad &\mbox{for $2 k_{\rm F} \le q \le 2 \pi - 4 k_{\rm F}$},\\
    N_5 (q/\pi), \quad &\mbox{for $2 \pi - 4 k_{\rm F} \le q \le \pi$},
   \end{array}\right.\label{Nqregion2}
\ee
and when $2/3 \le \bar{n} < 1$,
\be
  N(q) = \left\{
   \begin{array}{ll}
    N_1 (q/\pi), \quad &\mbox{for $0 \le q \le 2 \pi - 4 k_{\rm F}$},\\
    N_3 (q/\pi), \quad &\mbox{for $2 \pi - 4 k_{\rm F} \le q \le 2 k_{\rm F}$},\\
    N_5 (q/\pi), \quad &\mbox{for $2 k_{\rm F} \le q \le \pi$}.
   \end{array}\right.\label{Nqregion3}
\ee
It is clear that for $\bar{n} = 1/2$ and $2/3$ there are {\it two} different momentum regimes, otherwise 
there are {\it three} regimes.
For $0 < \bar{n} \le 1/2$ and $4 k_{\rm F} \le q \le \pi$, we find the following relation between $N(q)$ and 
$S(q)$: 
\be
N(q) + 4 S(q) = 2 \bar{n}.
\ee
The sum $N(q)+4 S(q)$ represents the density correlation between the same species of spin.~\cite{YO}
The fact that this quantity coincides with the noninteracting value $2 \bar{n}$ 
seems to be related to a hidden symmetry of the model.
Figure \ref{fig.IStjnq} shows the numerical results of $N(q)$ for 16 sites and various fillings.

Finally, we compare the results of $N(q,\omega)$ for the NN model with those for the IS model.
Figures \ref{fig.NNtjnqw}(a) and \ref{fig.NNtjnqw}(b) show the results for the 16-site NN {\it t-J} model with 
two holes and eight holes, respectively.
For two holes, charge dynamics for the NN interaction shares the global 
feature of the IS case.
However, there is a remarkable difference in contribution of higher-order spinons, holons, and 
antiholons. 
This corresponds to  [A] mentioned for $S(q,\omega)$ in the previous subsection.
For eight holes, the excitation spectrum with main intensity deviates strongly from the compact support for 
the IS interaction.  
This is due to modification of elementary excitations.~\cite{BBO}

\subsection{$A(k,\omega)$: electron removal from half-filling}

We examine the electron removal spectrum $A^- (k,\omega)$ from half-filling.
For the IS {\it t-J} model, excitation contents consist of one-spinon plus one-holon (1s1h) states and 
three-spinon plus one-holon (3s1h) states in the absence of antiholons.~\cite{HH}
Figure \ref{fig.IStjakwhf} shows the result of $A^- (k,\omega)$ for 16 sites.
Note that $\omega = E_0 (N_{\rm e}) - E_\nu (N_{\rm e}-1)$ is negative [see eq.\ (\ref{A-komega})].
The origin of energy is taken to be the ground-state energy at half-filling, and therefore the lowest energy 
at $k = \pi/2$ is equal to the chemical potential $\mu^- = E_0 ( N_{\rm e} ) - E_0 ( N_{\rm e} - 1 )$ for the 
finite-size system.
We notice that $\mu^-$ and $\mu^+$ given by $E_0 (N_{\rm e}+1) - E_0
(N_{\rm e})$ are discontinuous in the gapped system; in the present
case (i.e., Mott insulators with infinite energy for double
occupation) $\mu^+$ cannot be defined.
The region with finite intensity is compared with the support for the one-particle Green function in the 
high-density limit $\bar{n} \to 1$ (the solid lines of Fig.\ \ref{fig.IStjakwhf}).
In the thermodynamic limit, $\mu \, ( = \partial E_0 / \partial N_{\rm e} )$ becomes $- \pi^2 t/12$ (the 
broken 
line of Fig.\ \ref{fig.IStjakwhf}).
This value is consistent with $-0.822467t$ extrapolated from the numerical results for $N=4$, $8$, $12$ 
and $16$.
The excitation at $k=0$ has the energy $- \pi^2 t/3$, which agrees with the extrapolated value 
($-3.28987t$) from the numerical calculation for $N=4$-$16$.

Let us turn now to the magnitude of the intensity.
Recently, Kato has obtained the exact expression of the 1s1h contribution for $A^- (k,\omega)$.~\cite{Kato,Katocondmat}
For a finite system, the 1s1h contribution $G_{\rm 1s1h} (\tilde{x}, \tilde{t})$ to the hole propagator at 
half-filling is given by~\cite{Katocondmat}
\bea
  & & G_{\rm 1s1h} (\tilde{x}, \tilde{t}) = \frac{1}{2N} \nonumber \\
  & & + \frac{4 (-1)^{\tilde{x}}}{\pi N^2} \sum_{K=0}^{N/2-2} \; \sum_{I=0}^{N/2-K-2} \frac{\Gamma [ N/2 - K 
+ 1/2 ] \Gamma [ K + 1/2 ]}{\Gamma [ N/2 - K ] \Gamma [ K + 1 ]} \nonumber \\
  & & \times {\rm e}^{- {\rm i} \tilde{t} E_{IK}} \cos ( 2 \pi x ( I + K + 1 )/N ),
\eea
where $E_{IK} = 2 \pi^2 ( 2 I^2 - 2 I M_0 - 2 K^2 + 2 K M_0 + I + K - M_0 - 1/6 )/N^2 + \pi^2/3$ with $M_0 = 
N/2 - 1$ ($K$ and $I$ being integers), and $\Gamma [\cdots]$ is the gamma function.
From this result, we can read the momentum transfer $k$, the excitation energy $\omega$ from the 
half-filled ground state, and the form factor $M_{\rm 1s1h}$ as follows:
\bea
  & & k = 2 \pi ( I + K + 1 )/N - \pi, \qquad ({\rm mod} \; 2 \pi),\\
  & & \omega = E_{IK},\\
  & & M_{\rm 1s1h} = \frac{2}{\pi N} \cdot \frac{\Gamma [ N/2 - K + 1/2 ] \Gamma [ K + 1/2 ]}{\Gamma [ 
N/2 - K ] \Gamma [ K + 1 ]}.\quad
\eea
The 1s1h contribution for $N = 16$ is shown in Fig.\ \ref{fig.IStjakwhf2}(a).
It is found that at $k = \pi$, the 1s1h state does not contribute at all.~\cite{Kato}
\par

Analytical expression of the 3s1h contribution is not yet derived.
We can derive numerically the 3s1h contribution by subtracting the 1s1h contribution [Fig.\ 
\ref{fig.IStjakwhf2}(a)] from the total spectral weight [Fig.\
\ref{fig.IStjakwhf}].
We show the 3s1h contribution for $N = 16$ in Fig.\ \ref{fig.IStjakwhf2}(b).
Note that there are some cases where the 1s1h and 3s1h contributions
merge into a single pole; for example, $(k, \omega) = (\pi/4,
-3.43123t)$ and $(3 \pi/8, -4.89625t)$.
\par

Figure \ref{fig.IStjakwhf2}(b) gives us the following information about the 3s1h contribution:
\begin{itemize}
\item The 3s1h contribution is present even in the region of the $(k,\omega)$-plane where the 1s1h states 
dominate.
\item The 3s1h contribution has the peak intensity for each $k$ near the lower boundary of the 1s1h 
contribution.
This lower boundary for $0 \le k \le \pi$ corresponds to the spinon dispersion at half-filling.
\end{itemize}
\par

For each momentum, the pole positions of 3s1h states with dominant
intensity [the centers of black circles in Fig.\
\ref{fig.IStjakwhf2}(b)] find correspondence with the two-spinon
excitations in the HS model.
Namely, the values of $- (\omega - \omega_0)$ agree with the pole positions of $S(q,\omega)$ for the HS 
model, where $\omega_0$ denotes the position of the single pole at $k=0$.
However, those intensities are always weaker than the counterparts for the HS model.
These facts mean the following:
The 3s1h state with dominant intensity consists of two moving spinons
together with another spinon with zero energy and a holon with the maximal energy. 
The intensity is suppressed as compared with the two-spinon intensity,
namely $S(q,\omega)$ in the HS model, because of addition of a
further spinon and a holon.
Nevertheless the rough feature of the intensity distribution resembles $S(q,\omega)$ in the HS model.
\par

Next we study the spectrum of electron removal in the NN {\it t-J} model, for which the result is shown in 
Fig.\ \ref{fig.NNtjakwhf}.
Like the IS model, the number of poles for $k=0$ is only one.
The remarkable differences are as follows:

[a] There are many poles with small intensity outside the 3s1h continuum, although part of those poles are 
not apparent in the figure.

[b] The branch of the lowest energy for $0 \le k \le \pi$ is the cosine-like function, in contrast with the 
quadratic function for the IS model.

For the electron removal from half-filling, the elementary excitation
picture on the basis of spinons and holons applies
 only for the low-energy region in the NN supersymmetric {\it t-J} model.~\cite{BBO}
The difference between the NN case and the IS case appears in the dispersion relations of spinons and 
holons.
Taking account of the spin and charge conservations, 
possible excitations 
for $A^- (k,\omega)$ consist of the $N_{\rm sp}$-spinon plus one-holon states where $N_{\rm sp} = 1, 3, 5, 
\cdots, 
N-1$.
In the IS case the $N_{\rm sp} > 3$ contribution vanishes.

\subsection{$A(k,\omega)$: single-particle spectral function at less-than-half-filling}

As the single-particle excitation from the ground state for $\bar{n} < 1$, we consider the electron removal 
spectrum $A^- (k,\omega)$ and the electron addition one $A^+ (k,\omega)$.
The results for 16 sites with 14 electrons and 6 electrons are shown in Figs.\ \ref{fig.IStjakwlth}(a) and 
\ref{fig.IStjakwlth}(b), respectively.
The chemical potential for the finite-size system is defined as $\mu^- = E_0 (N_{\rm e}) - E_0 (N_{\rm e} - 1)
$ for the electron removal, and as $\mu^+ = E_0 (N_{\rm e} + 1) - E_0 (N_{\rm e})$ for the electron addition.
In the thermodynamic limit, $\mu^-$ and $\mu^+$ should be equal to $\mu$ given by
\be
  \mu / t = - \frac{1}{12} \pi^2 ( 3 \bar{n}^2 - 6 \bar{n} + 4 ).\label{cp}
\ee
This is obtained by differentiating the ground-state energy $E_0$~\cite{KY} with respect to $N_{\rm e}$.
The broken line of Fig.\ \ref{fig.IStjakwlth} indicates the value given by eq.\ (\ref{cp}).
The support by Ha and Haldane is shown by the solid lines.
Our result for $N = 16$ reproduces their support.

As the hole density becomes larger, the main intensity accumulates
along the edge of the compact support and across the chemical potential line.
This curve corresponds approximately to the energy band in the low-density limit $\bar{n} \to 0$.
The intensity of $A^+ (k,\omega)$ increases with increasing $\omega$ at each $k$.
On the other hand, the $\omega$-dependence of $A^- (k,\omega)$ is more intricate.
The complexity arises because more elementary excitations contribute to $A^- (k,\omega)$.
Namely, the excitation contents for $A^+ (k,\omega)$ consist of (s$_{\rm L}$, h$_{\rm L}$) + $\bar{{\rm h}}$ 
and the mirror states, while (s$_{\rm L}$, h$_{\rm L}$) + $\bar{{\rm h}}$ + 2(s$_{\rm R}$, h$_{\rm R}$) and the 
mirror states contribute to $A^- (k,\omega)$.~\cite{HH}

Now we consider the sum rule.
Because double occupation at each site is excluded in the {\it t-J} model, the following sum rule holds:
\be
  \int_{-\infty}^{\infty} {\rm d}\omega \left[ A^+ (k,\omega) + A^- (k,\omega) \right] = 1 - \frac{1}{2} \bar{n},
\label{sumrule}
\ee
independent of $k$ (see Appendix).
Integration of $A^- (k,\omega)$ over $\omega$ yields the momentum distribution for spin $\sigma$, which 
can be defined as,
\be
  n_{\sigma} (k) = \langle 0 | \tilde{c}^{\dagger}_{k \sigma} \tilde{c}_{k \sigma} | 0 \rangle,
\ee
where $| 0 \rangle$ denotes the ground state.
For the IS {\it t-J} model, the exact expression of $n_{\sigma} (k)$ is given by an infinite series.~\cite{MV}

Before closing this subsection, let us 
compare with the single-particle spectral functions of the NN {\it t-J} model.
In Figs.\ \ref{fig.NNtjakwlth}(a) and \ref{fig.NNtjakwlth}(b) we show the results of $A^{\pm} (k,\omega)$ for 16 
sites with two holes and ten holes, respectively.
The prominent difference from the IS case is the curvature of the branch with dominant intensity 
representing 
either a particle or a hole in the band picture.
Again, we find 
contributions of higher-order spinons, holons and antiholons in the high-energy region in the 
numerical data.
For $A^{\pm} (k,\omega)$ with two holes, the area of appreciable intensity is similar to the compact support 
in the corresponding IS case.
In $A^- (k,\omega)$ with ten holes, the region with main intensity shifts closer to the chemical 
potential~\cite{BBO} than that for the IS model.


\section{Summary and Discussion} \label{Summary}

We have investigated the dynamical properties of the 1D supersymmetric {\it t-J} model for the two types of 
the interaction.
The main results are summarized as follows:

(1) For the inverse-square type, the dynamical spin structure factor is independent of the electron density 
in the spectral region where only two spinons contribute.

(2) Concerning the electron removal spectrum from half-filling for the inverse-square type, the 
three-spinon plus one-holon states with dominant intensity have the correspondence with two-spinon 
excitations in the Haldane-Shastry model.

(3) For the nearest-neighbor type with lightly doped holes, the region with dominant intensity for a 
dynamical correlation function is similar to the corresponding compact support of the inverse-square type. 
However, the regions for the two types become different as the hole density is increased.

The feature (1) is an indication of the strong spin-charge separation in the zero-magnetic-field dynamics 
for the long-range supersymmetric {\it t-J} model.
In applied field, the strong spin-charge separation appears also in
the charge dynamics.~\cite{AYSK,Saiga}
We will report detailed results on the dynamics in nonzero magnetic field in a separate paper.
\par

The feature (3) suggests the following conjecture about the dynamics for the nearest-neighbor type:
In the thermodynamic limit, the most relevant states for a correlation
function would be 
composed of the excitation contents
(i.e., the minimal sets) by Ha and Haldane~\cite{HH}, provided the
electron density is near half-filling.
For large hole density, however, this  does not hold.
For the inverse-square type, the dynamics is characterized by 
spinons, holons and antiholons at arbitrary filling.
On the other hand, for the nearest-neighbor type, proper picture of
elementary excitations changes gradually with increasing hole density.
Namely the Ha-Haldane's excitation contents no longer apply 
even approximately to the case with a large hole density.
This is because the spinons and holons obeying the fractional
statistics interact with each other.
The charge carried by the holon and antiholon excitations in the
nearest-neighbor case is rather different from that in the
inverse-square case, even for quarter filling.~\cite{Ha}
\par

\section*{Acknowledgments}

We would acknowledge fruitful discussions with Y. Kato, T. Yamamoto and M. Arikawa.
The numerical calculations were performed partly at the Supercomputer Center of the Institute for Solid 
State Physics, University of Tokyo.
This work was supported by CREST from the Japan Science and Technology 
Corporation.

\appendix
\section{Sum Rule of $A(k,\omega)$}

In this appendix we prove eq.\ (\ref{sumrule}).
Integration of $A^+ (k,\omega) + A^- (k,\omega)$ over $\omega$ reduces to the quantity 
\be
  \langle 0|  \tilde{c}_{k \sigma} \tilde{c}^\dagger_{k \sigma}
  + \tilde{c}^\dagger_{k \sigma} \tilde{c}_{k \sigma}  |0
  \rangle.
\ee
Note that the tilde implies exclusion of double occupation at each 
site.
The Fourier transform:
\be
  \tilde{c}_{k \sigma} = \frac{1}{\sqrt{N}} \sum_j \tilde{c}_{j \sigma} {\rm
    e}^{- {\rm i} k j}, \quad \tilde{c}_{j \sigma} = c_{j \sigma} (1 - 
  n_{j,- \sigma}),
\ee
leads to
\bea
  \tilde{c}_{k \sigma} \tilde{c}^\dagger_{k \sigma}
  + \tilde{c}^\dagger_{k \sigma} \tilde{c}_{k \sigma} 
  &=& \frac{1}{N} \sum_{j,j'} ( c_{j \sigma} c^\dagger_{j' \sigma}
  + c^\dagger_{j' \sigma} c_{j \sigma} ) \nonumber \\
  & & \times (1 - n_{j,- \sigma}) (1 - n_{j',- \sigma})
  {\rm e}^{{\rm i} k (j'-j)} \nonumber \\
  &=& \frac{1}{N} \sum_j (1 - n_{j,- \sigma})^2 \nonumber \\
  &=& 1 - \frac{1}{N} \sum_j n_{j,- \sigma}.\label{ccdpcdc}
\eea
Here we have used the anticommutation relation of a fermionic operator
$c_{j \sigma}$.
The expectation value of the second term of eq.\ (\ref{ccdpcdc}) by
$|0 \rangle$ becomes $\bar{n}/2$ since the numbers of electrons with
$\sigma$ and $- \sigma$ are equal in the singlet ground state.
\par

This proof relies only on the condition that double occupation is
excluded at each site.
Thus eq.\ (\ref{sumrule}) holds irrespective of the interaction range
and the value of $J/t$ in the {\it t-J} model.
\par

\bigskip

\newpage


\begin{figure}
\caption{Enumeration of all possible ways to put an index on each box of the skew 
  Young diagram corresponding to the motif 01010.
The entries of the numbers 1, 2 and $\circ$ obey the rules [i] and [ii].
}
\label{fig.SYD}
\end{figure}

\begin{figure}
\caption{Skew Young diagrams corresponding to the motifs (a)
  01010$|$1111101$|$01100 and (b) 01010$|$1111110$|$11010.
In the case (a), there are other possibilities to put numbers or $\circ$ 
than the one shown in the left figure, while the right figure gives a unique combination in the case (b).
}
\label{fig.SYDab}
\end{figure}

\begin{figure}
\caption{$S(q,\omega)$ of the IS supersymmetric {\it t-J} model with 16 sites and (a) 2 holes and (b) 8 
holes.
The intensity of each pole is proportional to the {\it area} of the circle.
The solid lines show  dispersion relations of elementary excitations for
the compact support.
}
\label{fig.IStjsqw}
\end{figure}

\begin{figure}
\caption{$S(q)$ of the IS supersymmetric {\it t-J} model with 16 sites for various fillings.
}
\label{fig.IStjsq}
\end{figure}

\begin{figure}
\caption{$S(q,\omega)$ of the NN supersymmetric {\it t-J} model with 16 sites and (a) 2 holes and (b) 8 
holes.
The intensity of each pole is proportional to the {\it area} of the circle.
}
\label{fig.NNtjsqw}
\end{figure}

\begin{figure}
\caption{$N(q,\omega)$ of the IS supersymmetric {\it t-J} model with 16 sites and (a) 2 holes and (b) 8 
holes.
The intensity of each pole is proportional to the {\it area} of the circle.
The solid lines show dispersion relations of elementary excitations for
the compact support.
}
\label{fig.IStjnqw}
\end{figure}

\begin{figure}
\caption{$N(q)$ of the IS supersymmetric {\it t-J} model with 16 sites for various fillings.
}
\label{fig.IStjnq}
\end{figure}

\begin{figure}
\caption{$N(q,\omega)$ of the NN supersymmetric {\it t-J} model with 16 sites and (a) 2 holes and (b) 8 
holes.
The intensity of each pole is proportional to the {\it area} of the circle.
}
\label{fig.NNtjnqw}
\end{figure}

\begin{figure}
\caption{$A(k,\omega)$ of the IS supersymmetric {\it t-J} model with 16 sites at half-filling.
The intensity of each pole is proportional to the {\it area} of the circle.
The solid lines show  dispersion relations of elementary excitations for
the compact support.
The broken line indicates the chemical potential given by $- \pi^2 t/12$.
}
\label{fig.IStjakwhf}
\end{figure}

\begin{figure}
\caption{(a) The 1s1h contribution and (b) the 3s1h contribution to $A(k,\omega)$ of the IS supersymmetric 
{\it t-J} model with 16 sites at half-filling.
The intensity of each pole is proportional to the {\it area} of the circle.
The solid lines show  dispersion relations of elementary excitations for
the compact support.
The broken line indicates the chemical potential given by $- \pi^2 t/12$.
The meaning of black circles is explained in the text.
}
\label{fig.IStjakwhf2}
\end{figure}

\begin{figure}
\caption{$A(k,\omega)$ of the NN supersymmetric {\it t-J} model with 16 sites at half-filling.
The intensity of each pole is proportional to the {\it area} of the circle.
}
\label{fig.NNtjakwhf}
\end{figure}

\begin{figure}
\caption[]{$A(k,\omega)$ of the IS supersymmetric {\it t-J} model with 16 sites at (a) $\bar{n} = 7/8$ (14 
electrons) and (b) $\bar{n} = 3/8$ (6 electrons).
The intensity of each pole is proportional to the {\it area} of the circle.
The circles with dark and light shades correspond to $A^- (k,\omega)$ and $A^+ (k,\omega)$, respectively.
The solid lines show dispersion relations of elementary excitations for
the compact support.
The broken line indicates the chemical potential given by eq.\ (\ref{cp}).
}
\label{fig.IStjakwlth}
\end{figure}

\begin{figure}
\caption{$A(k,\omega)$ of the NN supersymmetric {\it t-J} model with 16 sites at (a) $\bar{n} = 7/8$ (14 
electrons) and (b) $\bar{n} = 3/8$ (6 electrons).
The intensity of each pole is proportional to the {\it area} of the circle.
The circles with dark and light shades correspond to $A^- (k,\omega)$ and $A^+ (k,\omega)$, respectively.
}
\label{fig.NNtjakwlth}
\end{figure}


\begin{thebibliography}{99}

\bibitem{Ochiai} A. Ochiai, T. Suzuki and T. Kasuya:
J. Phys. Soc. Jpn. {\bf 59} (1990) 4129.

\bibitem{Kohgi97} M. Kohgi, K. Iwasa, J. -M. Mignot, A. Ochiai and T. Suzuki:
Phys. Rev. B {\bf 56} (1997) R11388.

\bibitem{Kohgi99} M. Kohgi, K. Iwasa, J. -M. Mignot, N. Pyka, H. Kadowaki, A. Ochiai, H. Aoki and T. Suzuki:
Physics of Strongly Correlated Electron Systems, JJAP Series 11, (ed.
T. Komatsubara {\it et al.}, Publication Office, Japanese Journal of Applied
Physics, 1999), pp.111.

\bibitem{Carter} S. A. Carter, B. Batlogg, R. J. Cava, J. J.
Krajewski, W. F. Peck, Jr. and T. M. Rice:
Phys. Rev. Lett. {\bf 77} (1996) 1378.

\bibitem{Kim96} C. Kim, A. Y. Matsuura, Z. -X. Shen, N. Motoyama, H. Eisaki, S. Uchida, T. Tohyama and S. 
Maekawa: 
Phys. Rev. Lett. {\bf 77} (1996) 4054.

\bibitem{Kim97} C. Kim, Z. -X. Shen, N. Motoyama, H. Eisaki, S. Uchida, T. Tohyama and S. Maekawa: 
Phys. Rev. B {\bf 56} (1997) 15589.

\bibitem{Fujisawa} H. Fujisawa, T. Yokoya, T. Takahashi, S. Miyasaka, M. Kibune and H. Takagi: 
Solid State Commun. {\bf 106} (1998) 543.

\bibitem{Mizokawa} T. Mizokawa, C. Kim, Z. -X. Shen, A. Ino, A. Fujimori, M. Goto, H. Eisaki, S. Uchida, M. 
Tagami, K. Yoshida, A. I. Rykov, Y. Siobara and S. Tajima: 
preprint.



\bibitem{THM} T. Tohyama, P. Horsch and S. Maekawa:
Phys. Rev. Lett. {\bf 74} (1995) 980.

\bibitem{ZKMS} S. Zhang, M. Karbach, G. M\"uller and J. Stolze:
Phys. Rev. B {\bf 55} (1997) 6491.

\bibitem{SP} S. Sorella and A. Parola: 
J. Phys. Condens. Matter {\bf 4} (1992) 3589.

\bibitem{MTY} S. Maekawa, T. Tohyama and S. Yunoki:
Physica C {\bf 263} (1996) 61.

\bibitem{TM} T. Tohyama and S. Maekawa:
J. Phys. Soc. Jpn. {\bf 65} (1996) 1902.

\bibitem{PHMS96} K. Penc, K. Hallberg, F. Mila and H. Shiba: 
Phys. Rev. Lett. {\bf 77} (1996) 1390.

\bibitem{PHMS97} K. Penc, K. Hallberg, F. Mila and H. Shiba: 
Phys. Rev. B {\bf 55} (1997) 15475.

\bibitem{FHPMD} J. Favand, S. Haas, K. Penc, F. Mila and E. Dagotto: 
Phys. Rev. B {\bf 55} (1997) R4859.

\bibitem{EO} R. Eder and Y. Ohta: 
Phys. Rev. B {\bf 56} (1997) 2542.



\bibitem{KY} Y. Kuramoto and H. Yokoyama:
Phys. Rev. Lett. {\bf 67} (1991) 1338.

\bibitem{Haldane91} F. D. M. Haldane:
Phys. Rev. Lett. {\bf 67} (1991) 937.

\bibitem{HH} Z. N. C. Ha and F. D. M. Haldane:
Phys. Rev. Lett. {\bf 73} (1994) 2887.

\bibitem{Kato} Y. Kato:
Phys. Rev. Lett. {\bf 81} (1998) 5402.

\bibitem{AYSK} M. Arikawa, T. Yamamoto, Y. Saiga and Y. Kuramoto: 
to be published.

\bibitem{SutherlandtJ} B. Sutherland: 
Phys. Rev. B {\bf 12} (1975) 3795.



\bibitem{Sutherland} B. Sutherland: 
Phys. Rev. A {\bf 4} (1971) 2019.

\bibitem{FK} T. Fukui and N. Kawakami: 
Phys. Rev. B {\bf 54} (1996) 5346. 

\bibitem{OLSA} M. Ogata, M. U. Luchini, S. Sorella and F. F. Assaad: 
Phys. Rev. Lett. {\bf 66} (1991) 2388. 

\bibitem{LW} J. T. Liu and D. F. Wang: 
Phys. Rev. B {\bf 56} (1997) 2312. 

\bibitem{GB} E. R. Gagliano and C. A. Balseiro: 
Phys. Rev. Lett. {\bf 59} (1987) 2999. 

\bibitem{comment} We can know the numerical accuracy in the special case where an exact result is 
obtained in the finite system; for example, the one-spinon plus one-holon contribution to $A^- (k,\omega)$ 
from half-filling (see $\S$3.5).




\bibitem{Kawakami} N. Kawakami: 
Phys. Rev. B {\bf 45} (1992) 7525. 

\bibitem{WLC} D. F. Wang, J. T. Liu and P. Coleman: 
Phys. Rev. B {\bf 46} (1992) 6639. 

\bibitem{KKN} A. N. Kirillov, A. Kuniba and T. Nakanishi:
Commun. Math. Phys. {\bf 185} (1997) 441.

\bibitem{HB} K. Hikami and B. Basu-Mallick:
preprint (math-ph/9904033).

\bibitem{Haldane88} F. D. M. Haldane: 
Phys. Rev. Lett. {\bf 60} (1988) 635. 

\bibitem{Shastry88} B. S. Shastry:
Phys. Rev. Lett. {\bf 60} (1988) 639. 

\bibitem{KK} Y. Kato and Y. Kuramoto: 
J. Phys. Soc. Jpn. {\bf 65} (1996) 1622.

\bibitem{HZ} F. D. M. Haldane and M. R. Zirnbauer: 
Phys. Rev. Lett. {\bf 71} (1993) 4055. 

\bibitem{GV} F. Gebhard and D. Vollhardt: 
Phys. Rev. B {\bf 38} (1988) 6911. 

\bibitem{Gutzwiller} M. C. Gutzwiller: 
Phys. Rev. Lett. {\bf 10} (1963) 159. 

\bibitem{Schlottmann} P. Schlottmann: 
J. Phys. Condens. Matter {\bf 5} (1993) 313. 

\bibitem{BBO} P. -A. Bares, G. Blatter and M. Ogata: 
Phys. Rev. B {\bf 44} (1991) 130. 

\bibitem{YO} H. Yokoyama and M. Ogata: 
Phys. Rev. B {\bf 53} (1996) 5758. 

\bibitem{Katocondmat} Y. Kato:
preprint (cond-mat/9806202).

\bibitem{MV} W. Metzner and D. Vollhardt: 
Phys. Rev. B {\bf 37} (1988) 7382. 

\bibitem{Saiga} Y. Saiga: 
Thesis (Tohoku University, 1999).

\bibitem{Ha} Z. N. C. Ha:
{\it Quantum Many-Body Systems in One Dimension} (World Scientific,
Singapore, 1996).

\end{thebibliography}
\end{document}